\documentstyle[12pt]{article}
\newcommand{\cerchio}{\mbox{\begin{picture}(2,2)(0,0)\put(1,1){\circle{2}}\end{picture}}}

\newcommand{\Tr}{\mbox{Tr }}
\renewcommand{\Pr}{\hat{\mbox{I}\!\!\mbox{P}}}
\newcommand{\eff}{\mbox{\small eff}}
\newcommand{\latt}{\mbox{ latt}}
\newcommand{\DR}{{\rm DR}}
\newcommand{\DRED}{{\rm DRED}}
\newcommand{\NDR}{{\rm  NDR}}

\newcommand{\id}{\mbox{1$\!\!$I}}
\newcommand{\sid}{\mbox{\small 1$\!\!$I}}

\newcommand{\nn}{\nonumber}
\newcommand{\<}{\langle}
\renewcommand{\>}{\rangle}

\begin{document}

\pagestyle{empty}
\begin{flushright}
CERN-TH/95-23 \\
ROME prep. 1084/95\\
SHEP 95-25 \\
\end{flushright}
\vskip 0.5cm
\centerline{\bf NON-PERTURBATIVE RENORMALISATION OF THE}
\centerline{\bf LATTICE $\Delta S=2$ FOUR-FERMION OPERATOR}
\vskip 0.8cm
\centerline{\bf A. Donini$^a$, G. Martinelli$^{b,*}$, C.T. Sachrajda$^c$,
M. Talevi$^a$ and A. Vladikas$^{d}$ }
\centerline{$^a$ Dip. di Fisica,
Universit\`a degli Studi di Roma ``La Sapienza'' and}
\centerline{INFN, Sezione di Roma, P.le A. Moro 2, 00185 Rome, Italy. }
\centerline{$^b$ Theory Division, CERN, 1211 Geneva 23, Switzerland.}
\centerline{$^c$ Dep. of Physics, University of Southampton,}
\centerline{Southampton SO17 1BJ, U.K.}
\centerline{$^d$ INFN, Sezione di Roma II and}
\centerline{Dip. di Fisica, Universit\`a di Roma ``Tor Vergata'',}
\centerline{Via della Ricerca Scientifica 1, 00133 Rome, Italy.}

\begin{abstract}
We compute the renormalised four-fermion operator $O^{\Delta S=2}$ using  a
non-perturbative method recently introduced for determining the
renormalisation constants of generic lattice composite operators.
Because of the presence of  the Wilson term,
$O^{\Delta S=2}$ mixes with operators of different chiralities.
A projection method to determine the mixing coefficients is implemented.
The numerical results for the renormalisation constants have been obtained
from a simulation performed using the
SW-Clover quark action, on a $16^3 \times 32$ lattice, at $\beta=6.0$.
We show that the use of the  constants determined non-perturbatively
 improves  the chiral behaviour of the
lattice kaon matrix element $\<\bar K^0| O^{\Delta S=2} | K^0\>_{\latt}$.
\end{abstract}
\vskip 0.5cm

\begin{flushleft}
CERN-TH/95-23 \\
August 1995
\end{flushleft}
\vskip 0.5 cm
\centerline{$^*$ On leave of absence from Dip. di Fisica,
Universit\`a degli Studi ``La Sapienza'', Rome, Italy.}

\newpage
\pagestyle{plain}
\setcounter{page}{1}

\section{Introduction}
\label{sec:intro}

Renormalisation of lattice operators is a necessary step for  obtaining
physical
results from numerical simulations.  In this paper, we apply
the general method introduced in \cite{NP} to the four-fermion
operator\footnote{We use the Euclidean metric  throughout this
paper.} \begin{equation}
O^{\Delta S=2}=(\bar s \gamma_{\mu}^L d )(\bar s \gamma_{\mu}^L d)
\, , \label{eq:O_DS=2} \end{equation}
 which appears in  the weak effective Hamiltonian relevant
for $K^0$--$\bar K^0$ mixing
\begin{equation}
{\cal H}^{\Delta S=2}_{\eff}=C(M_W/\mu) O^{\Delta S=2}(\mu) \, ,
\end{equation}
where $\gamma_{\mu}^L=\frac{1}{2}\gamma_{\mu} (1-\gamma_5)$,
$O^{\Delta S=2}(\mu)$ is the renormalised operator,
 $C(M_W/\mu)$  is
the corresponding Wilson coefficient and $\mu$ the renormalisation scale.
The $K^0$--$\bar K^0$  matrix element of $O^{\Delta S=2}(\mu) $
defines  the so-called kaon  $B$-parameter
\begin{equation}
\<\bar K^0| O^{\Delta S=2}(\mu) | K^0\>=\frac{8}{3}f_K^2m_K^2B_K(\mu)\, .
 \label{eq:B_K}
\end{equation}
The uncertainty in the value of this matrix element restricts the
precision with which the CKM matrix elements $\rho$ and $\eta$ (in
the Wolfenstein parametrisation) can be determined from experimental
measurements. It is therefore of considerable importance to determine
this matrix element using lattice simulations.

In the continuum, chiral symmetry implies that
the kaon matrix element of $O^{\Delta S=2}$ vanishes in the chiral
limit \cite{Cabibbo,Gellmann}
\begin{equation}
\<\bar K^0(q) | O^{\Delta S=2}(\mu) | K^0(p)\>=
                \gamma (p\cdot q) + O\left(  (p\cdot q)^2  \right) \, .
\label{eq:B_K_chiral}  \end{equation}
On the lattice however, in simulations based on Wilson's formulation of
the fermion action
(such as the standard Wilson action or the SW-Clover action \cite{sw}),
the presence of chiral symmetry breaking terms leads to the mixing of
$O^{\Delta S=2}$ with operators of different chirality
\cite{marti84}--\cite{improved}, and the matrix element of  $O^{\Delta
S=2}$ is different from zero at $p \cdot q=0$
\cite{bern}--\cite{Crisafulli}. For this reason, it is possible to
define a renormalised operator with definite chiral properties only in
the continuum limit, i.e. when $a \to 0$. At finite $a$, one can
improve the chiral behaviour of the matrix element of $O^{\Delta S=2}$,
by sub\-tracting a suitable set of dimension six operators. The mixing
coefficients have so far been computed  only in one-loop perturbation
theory \cite{marti84},\cite{berw}--\cite{improved}. In this way, the
systematic error in the value of the matrix element determined on the
lattice  is of $O(\alpha_s^2)$.  In addition, as a consequence of the
finiteness of the lattice spacing, there are errors of $O(a)$.
Following Symanzik's proposal, one can reduce these discretization
errors from $O(a)$ to $O(\alpha_s a)$ by using the   tree-level
``improved''  SW-Clover lattice quark action  \cite{sw,clover}.   Using
this action, the  improvement has been shown to be effective
for two-fermion operators, at values of $\beta$ currently used in
numerical simulations \cite{msv}--\cite{wiukqcd}. It remains true
however, that ignorance of higher-order perturbative corrections to the
mixing coefficients can distort the chiral behaviour of the operator
and hence induce a large systematic error in the determination of
$B_K$.  The use of a non-perturbative approach to the determination of
the renormalisation constants, should reduce this systematic effect.

In the following, we will define a renormalised operator $O^{\Delta
S=2}(\mu)$, obtained by applying the non-perturbative method of ref.
\cite{NP} to the computation of the mixing coefficients and of the
overall renormalisation constant.
In order to reduce the discretisation errors, including those induced
by the mixing with higher dimensional operators, it is necessary to
use an improved fermion action and operators. In the computations
described below we have used the improved SW-Clover action and the
``improved-improved'' operators introduced in ref.
\cite{improved,tass2}.  We  monitor the effects of the
non-perturbative determination of the mixing coefficients by comparing
the chiral behaviour  of the matrix element $\<\bar K^0(q)|O^{\Delta
S=2}|K^0(p)\>_{\latt}$ computed by using  the operator renormalised
with  standard or boosted \cite{Lepage}  perturbation theory to  the
matrix  element of the operator  renormalised non-perturbatively. In
particular, by parametrizing the matrix element  near the  chiral limit
in the standard way \cite{capri}--\cite{gupta},
\begin{equation}
\<\bar K^0|O^{\Delta S=2}|K^0\>_{\latt}=\alpha+\beta m_K^2+\gamma (p\cdot q)
       +\delta m_K^4+\epsilon m_K^2(p\cdot q) +\zeta (p\cdot q)^2+\ldots ,
\label{eq:B_K_latt}\end{equation}
we investigate the differences in the values of
$\alpha$ and $\beta$, obtained by fitting the dependence of the
matrix element on the kaon masses and momenta.
Since in the continuum
$\alpha$ and $\beta$ are absent, cf. eq.\ (\ref{eq:B_K_chiral}), we consider a
reduction of their values  as a measure of the improvement in the chiral
behaviour and in the accuracy of the determination of the matrix element.
Using the data of the APE collaboration  \cite{Donini94,Crisafulli},
we show that the chiral behaviour is indeed improved by
using the non-perturbative results.

The paper is organized as follows. In section \ref{sec:strategy}, we
briefly summarize the strategy followed for computing the mixing
coefficients and the overall renormalisation constant of the relevant
four-fermion operator; in section \ref{sec:mixing}, we illustrate the
projection method used to determine the mixing coefficients.
Although the method is applied specifically to the renormalisation
of the operator $O^{\Delta S=2}$, it can readily be generalised
to other sets of operators which mix under renormalisation.
In section~\ref{sec:PT} we give some information about the perturbative
evaluation of the renormalisation constants on the lattice;
in section~\ref{sec:numerical} we present the details of the numerical
simulation and discuss our results and, finally, we present our conclusions
in section~\ref{sec:conclusion}.

\section{The non-perturbative method for four-fer\-mion operators}
\label{sec:strategy}
The renormalisation method proposed in \cite{NP} completely avoids  the
use of lattice perturbation theory and allows for a non-perturbative
determination of the renormalisation constants of any composite
operator in a renormalisation scheme which is independent of the method
used to regulate the ultra-violet divergences.   In particular, the
renormalised operators are independent of the fact that we start from
bare operators in lattice QCD. To stress this point further we will
refer to the renormalisation scheme defined below for $O^{\Delta S=2}$
as the RI (Regularization Independent) scheme \cite{Ciuchini2}\footnote{
 Although of course such
a name could be applied equally well to many other schemes.}.
Non-perturbative renormalisation conditions are imposed directly on
quark Green functions with off-shell external states  in a fixed gauge,
for example the Landau gauge. The method is expected to work in all
cases where it is possible to fix the virtuality of the external states
$p^2=\mu^2$ so as to satisfy the condition $\Lambda_{{\rm QCD}}\ll \mu \ll
1/a$. The condition $\mu \gg \Lambda_{{\rm QCD}}$ is necessary because one
has to match perturbatively the effective Hamiltonian,  expressed in
terms of operators renormalised at the scale $\mu$, to the full theory.
This condition is common to all approaches currently used.  The
condition $\mu \ll 1/a$ is a requirement common to all lattice methods
and is due to the presence of $O(a)$ ($O(\alpha_s a)$) effects in the
operator matrix elements. The existence of the ``window"
$\Lambda_{{\rm QCD}}\ll \mu \ll 1/a$ depends on  the value of the bare
lattice coupling $\beta$ at which the numerical calculations are
performed. We refer the reader to ref. \cite{NP} for a more detailed
discussion on this point.

In the following, in order to use a more transparent notation in the formulae,
we will consider the operator
\begin{equation}
O_+= \frac{1}{2}
[(\bar\psi_1\gamma^L_{\mu}\psi_2)(\bar\psi_3\gamma^L_{\mu}\psi_4)
                 +(2\leftrightarrow 4)],
\label{eq:O_+} \end{equation}
with four distinct quark flavours ($f=1,2,3,4$) instead of the operator
$O^{\Delta S=2}$. $O_+$ and $O^{\Delta S=2}$ have the same
renormalisation properties.

The discretization of the quark action {\em \`a la} Wilson, induces a
mixing of the operator (\ref{eq:O_+}) with operators of a different
chirality  which, in the language of refs.
\cite{Ciuchini2}--\cite{Ciuchini}, correspond to the so-called
``effervescent''  (``evanescent'') operators.   The mixing, being a
consequence of the regularization procedure,  is not limited to the
lattice case, but is present also in continuum regularizations. The
effervescent operators must be subtracted from the bare one by a
suitable renormalisation procedure \footnote{ Using dimensional
regularization, the one-loop mixing with the ``effervescent''
operators is cancelled by the minimal subtraction of the pole in
$1/\epsilon$.}.

In the lattice case, CPS symmetry
fixes the basis of operators that may
appear in perturbation theory  \cite{BERNARD2}:
\begin{eqnarray}
O^{SP}_+&=&-\frac{1}{16N_c}
        [(\bar\psi_1\psi_2)(\bar\psi_3\psi_4)
        -(\bar\psi_1\gamma_5\psi_2)(\bar\psi_3\gamma_5\psi_4)
         +(2\leftrightarrow 4) ],                    \label{eq:O_+^SP} \\
O^{VA}_+&=&-\frac{(N_c^2+N_c-1)}{32N_c}
  [(\bar\psi_1\gamma_{\mu}\psi_2)(\bar\psi_3\gamma_{\mu}\psi_4)
-(\bar\psi_1\gamma_{\mu}\gamma_5\psi_2)(\bar\psi_3\gamma_{\mu}\gamma_5\psi_4)
                                                       \nn \\
        & &\qquad     +(2\leftrightarrow 4) ],  \label{eq:O_+^VA} \\
O^{SPT}_+&=&\frac{(N_c-1)}{16N_c}
       [(\bar\psi_1\psi_2)(\bar\psi_3\psi_4)
       +(\bar\psi_1\gamma_5\psi_2)(\bar\psi_3\gamma_5\psi_4) \nn \\
       & & \qquad
       +(\bar\psi_1\sigma_{\mu\nu}\psi_2)(\bar\psi_3\sigma_{\mu\nu}\psi_4)
       +(2\leftrightarrow 4) ],                     \label{eq:O_+^SPT}
\end{eqnarray}
where $N_c=3$ denotes the number of colours.

We renormalise the operator $O_+$ by introducing the subtracted
operator $O^s_+$,
\begin{equation}
O_+(\mu)= Z_+ O^s_+=Z_+
(O_+ +Z_1 O^{SP}_+ +Z_2 O^{VA}_+ +Z_3 O^{SPT}_+)\, ,
\label{eq:O_+(mu)}\end{equation}
where $Z_{+,1,2,3}=Z_{+,1,2,3}(\mu a , g_0^2(a))$ and the bare lattice
coupling is given by the relation $\beta= 6/ g_0^2(a)$. The mixing
constants $Z_i,\ i=1,\ldots,3$, are determined by means of projection
operators that will be defined in sec\-tion~\ref{sec:mixing}. Their
values are fixed by the requirement that, up to terms of $O(\alpha_s
a)$, $O^s_+$ renormalises multiplicatively.  $O^s_+$ is logarithmically
divergent as $a\to  0$, and this divergence is removed by imposing a
renormalisation condition on $O^s_+$ which defines  the overall
renormalisation constant $Z_+(\mu a,g^2_0(a))$,
\begin{equation}
Z_+(\mu a,g^2_0(a))Z_{\psi}^{-2}(\mu a,g^2_0(a))
\Gamma^s_+(pa)|_{p^2=\mu^2}=1,
\label{eq:Z_+}\end{equation}
where $\Gamma^s_+(pa)$ is obtained by projecting a suitable amputated
Green  function of the operator $O^s_+$  on the  Dirac structure
$\gamma^L_{\mu}\otimes \gamma^L_{\mu}$ (see eq.\  (\ref{eq:P_0}) in
section \ref{sec:mixing} and refs.\ \cite{NP,Ciuchini2}). $Z_{\psi}$ is
the quark field renormalisation constant to be defined below (eq.\
(\ref{eq:Z_psi}) of section  \ref{sec:mixing} and ref. \cite{NP}). In
eq.\ (\ref{eq:Z_+}), $p^2=\mu^2$ denotes the momentum of the external
quark states. We have chosen equal momenta for all four external quark
legs, because this is the simplest choice which regulates the infrared
divergences \cite{Ciuchini2}.

The renormalised operator in eq.\ (\ref{eq:O_+(mu)}), calculated in the
RI scheme, depends both on the gauge and on the external states.  The
Wilson coefficient must also be calculated in the same gauge and with
the same external states in order to obtain the physical operators
which are independent of both\footnote{This is true up to higher order
continuum perturbative corrections and lattice systematic errors.}.
The next-to-leading order calculation of the Wilson coefficient
relevant for the operator (\ref{eq:O_+}), in the Landau gauge and with
equal external momenta, can be found in ref. \cite{Ciuchini2}.

\section{Determination of the mixing constants}
\label{sec:mixing}
In this section, we define the four-point amputated Green functions and
introduce the projectors that have been used to determine the mixing
constants.

Since the non-perturbative renormalisation conditions are imposed  on
quark states, the Green functions of a four-fermion operator will
depend on four coordinates. Denoting by $x_1,x_3$ and $x_2,x_4$  the
coordinates of the outgoing and incoming quarks, the Green functions
corresponding to the insertion of  the operators
(\ref{eq:O_+})--(\ref{eq:O_+^SPT}) can be written as linear
combinations of Green functions of the form
\begin{equation}
G_{\Gamma^a}(x_1,x_2,x_3,x_4)=\<\psi_1(x_1)\bar\psi_2(x_2) O_{\Gamma^a}(0)
                      \psi_3(x_3)\bar\psi_4(x_4)\>\, ,  \label{eq:G_Gamma(x)}
\end{equation}
where $\<\cdots\>$ denotes the  vacuum expectation value, i.e. the
average over the gauge-field configurations.
The generic four-fermion operator $O_{\Gamma^a}$ is given by
\begin{equation}
O_{\Gamma^a}(0)=C_{\Gamma^a}\left[
    \bar\psi_1(0)\Gamma^a\psi_2(0) \bar\psi_3(0)\Gamma^a\psi_4(0)
   +\bar\psi_1(0)\Gamma^a\psi_4(0) \bar\psi_3(0)\Gamma^a\psi_2(0)
\right]\, ,
\label{eq:O_Gamma}\end{equation}
where $\Gamma^a$ denotes a Dirac matrix, and $C_{\Gamma^a}$ is a
constant associated with $\Gamma^a$. The index $a$ can be either
single-valued (if $\Gamma^a = \id$ or $\gamma_5$)  or be summed
over a range of values (if $\Gamma^a = \gamma_{\mu}$,
$\gamma_{\mu}\gamma_5$ or $\sigma_{\mu\nu}$ a sum over repeated Lorentz
indices is implied).

The Fourier transform of the non-amputated Green function
(\ref{eq:G_Gamma(x)}), at equal external momenta $p$, has the form
\begin{equation}
G_{\Gamma^a}(p)^{ABCD}_{\alpha\beta\gamma\delta}
  =C_{\Gamma^a}\left[
  \<\Gamma^a(p)^{AB}_{\alpha\beta}\otimes \Gamma^a(p)^{CD}_{\gamma\delta}\>
 -\<\Gamma^a(p)^{AD}_{\alpha\delta}\otimes\Gamma^a(p)^{CB}_{\gamma\beta}\>
\right]\, ,
\label{eq:G_Gamma(p)}\end{equation}
where
\begin{equation}
\Gamma^a(p)^{XY}_{\chi\psi}=S(p|0)^{XR}_{\chi\rho}\Gamma^a_{\rho\sigma}
                   (\gamma_5S(p|0)^{\dag}\gamma_5)^{RY}_{\sigma\psi}\, .
\label{eq:gammaa} \end{equation}
In eqs. (\ref{eq:G_Gamma(p)}) and (\ref{eq:gammaa}) the upper-case
Roman superscripts denote colour labels and the lower case Greek
subscripts denote spinor labels. $S(p|0)$ is defined by
\begin{equation}
S(p|0)=\int d^4x S(x|0) e^{-ip\cdot x},
\end{equation}
where $S(x|0)$ is the quark propagator computed on a single gauge-field
configuration (cf. section 4 of \cite{NP}), and is therefore not
translationally invariant. It satisfies the relation
\begin{equation}
S(x|0)=\gamma_5 S^{\dag}(0|x)\gamma_5.
\end{equation}
The amputated Green function can be obtained from  eq.\
(\ref{eq:G_Gamma(p)})
\begin{equation}
\Lambda_{\Gamma^a}(p)^{RSR'S'}_{\rho\sigma\rho'\sigma'}
=S^{-1}(p)^{RA}_{\rho\alpha}S^{-1}(p)^{R'C}_{\rho'\gamma}
 G_{\Gamma^a}(p)^{ABCD}_{\alpha\beta\gamma\delta}
 S^{-1}(p)^{BS}_{\beta\sigma}S^{-1}(p)^{DS'}_{\delta\sigma'}\, ,
\label{eq:Lambda_Gamma(p)}\end{equation}
where $S(p)$ is the Fourier transform of the translationally-invariant
quark propagator, i.e. the Fourier transform of $S(x|0)$, averaged over
the gauge-field configurations.

As mentioned above, the renormalisation procedure necessary to
determine the mixing constants  consists in defining suitable
projectors on the amputated Green functions  of the operators
(\ref{eq:O_+})--(\ref{eq:O_+^SPT}). To this end let us introduce a more
convenient notation. Let us denote by $O_i,\ i=0,\ldots,3$,
respectively, the operators $O_+,O_+^{SP},O_+^{VA},O_+^{SPT}$. Then,
the projectors $\Pr_i,\ i=0,\ldots,3$, are defined by the condition
\begin{equation}
\Tr \Pr_i \Lambda^{(0)}_j=\delta_{ij},\qquad i,j=0,\ldots,3, \label{eq:P_i}
\end{equation}
where $\Lambda^{(0)}_i,\ i=0,\ldots,3$, are the amputated Green
functions, at tree level, of the operators $O_i$, and the trace is
understood over colour and spin (as defined below).
The renormalisation scheme depends on the precise definition of the
projection operators and we now define our procedure in detail.

For each Dirac Matrix $\Gamma^b$, we define the projector
$\Pr_{\Gamma^b}$ by~\footnote{It is only the traces
(\ref{eq:proj_def_1}) which are required for the determination of the
subtraction constants.}
\begin{eqnarray}
\Tr\Pr_{\Gamma^b} \Lambda_{\Gamma^a}(p)
=(\Gamma^b_{\sigma\rho}\otimes \Gamma^b_{\sigma'\rho'})
 \Lambda_{\Gamma^a}(p)^{RRR'R'}_{\rho\sigma\rho'\sigma'}\, ,
\label{eq:proj_def_1}
\end{eqnarray}
where the index $b$ is either fixed or corresponds to a sum over
repeated indices. In the free theory, the amputated Green function
reduces to
\begin{equation}
\Lambda^{(0)}_{\Gamma^a}(p)^{RSR'S'}_{\rho\sigma\rho'\sigma'}
=C_{\Gamma^a}[
 \delta^{RS}\delta^{R'S'}(\Gamma^a_{\rho\sigma}\otimes\Gamma^a_{\rho'\sigma'})
-\delta^{RS'}\delta^{R'S}(\Gamma^a_{\rho\sigma'}\otimes\Gamma^a_{\rho'\sigma})].
\label{eq:Lambda_+_0}\end{equation}
and the result of the projection defined in (\ref{eq:proj_def_1}) is:
\begin{eqnarray}
\Tr \Pr_{\Gamma^b} \Lambda^{(0)}_{\Gamma^a}(p)
=C_{\Gamma^a}[
    N_c^2 (\Tr \Gamma^a \Gamma^b)(\Tr \Gamma^a \Gamma^b)
  - N_c   (\Tr \Gamma^a\Gamma^b\Gamma^a\Gamma^b)].
\label{eq:proj_free_1}\end{eqnarray}
The projectors corresponding to  $O_+$ and to the operators $O_i$ defined in
eqs.(\ref{eq:O_+^SP})--(\ref{eq:O_+^SPT}) are as follows:
\begin{eqnarray}
\Pr_0&=&\frac{1}{8N_c(N_c+1)}\Pr_{\gamma^R_{\mu}} ,          \label{eq:P_0}\\
\Pr_1&=&\frac{N_c}{2(1-N_c^2)}(\Pr_{\sid}-\Pr_{\gamma_5}) \nn \\
     &+&\frac{1}{4(1-N_c^2)}(\Pr_{\gamma_{\mu}}-\Pr_{\gamma_{\mu}\gamma_5}),
                                                             \label{eq:P_1}\\
\Pr_2&=&\frac{1}{2(1-N_c^2)(N_c^2+N_c-1)}(\Pr_{\sid}-\Pr_{\gamma_5}) \nn \\
     &+&\frac{N_c}{4(1-N_c^2)(N_c^2+N_c-1)}
        (\Pr_{\gamma_{\mu}}-\Pr_{\gamma_{\mu}\gamma_5})      \label{eq:P_2}\\
\Pr_3&=&\frac{1}{8(N_c^2-1)}(\Pr_{\sid}+\Pr_{\gamma_5}+\Pr_{\sigma_{\mu\nu}}),
\label{eq:P_3}\end{eqnarray}
where $\gamma^R_{\mu}=\frac{1}{2}\gamma_{\mu}(1-\gamma_5)$. Note that
the projectors $\Pr_0$ and $\Pr_3$, eqs. (\ref{eq:P_0}) and
(\ref{eq:P_3}), have the same Dirac structure as the  operators $O_0$
and $O_3$, eqs. (\ref{eq:O_+}) and (\ref{eq:O_+^SPT}). This is due to
the Fierz rearrangement properties of these operators.

It is possible to determine the mixing coefficients
$Z_i$  by using the projectors (\ref{eq:P_0})--(\ref{eq:P_3}) defined in
the free field case. Let us introduce the  matrix  $D$ defined by
\begin{equation} \Lambda_i=\sum_{j=0}^3 D_{ij} \Lambda^{(0)}_j
\, , \end{equation}
where the elements $D_{ik}$ are determined non-perturbatively by the
projections
\begin{equation}
D_{ik}=\Tr \Pr_k \Lambda_i,\qquad i,k=0,\ldots,3.
\end{equation}
The mixing constants $Z_i$ are then fixed  by the condition that the
subtracted operator $O_+^s$ is proportional to the bare free
operator
\begin{equation}
\Tr \Pr_k \Lambda^s_+=
    \left(D_{0k}+\sum_{i=1}^3Z_i D_{ik} \right)=0,\qquad k=1,2,3\, .
\label{eq:mixing_condition}
\end{equation}
Equation (\ref{eq:mixing_condition}) yields three conditions
corresponding to a linear non-homo\-geneous system in the three unknowns
$Z_i$.   Defining the reduced $3\times 3$ matrix $\tilde D$ as
\begin{equation}
\tilde D_{ik}=D_{ki},\qquad i,k=1,2,3 \, ,
\end{equation}
the solutions of this linear system are given by
\begin{equation}
Z_i=-\sum_{k=1}^3(\tilde D)^{-1}_{ik}D_{0k},\qquad i=1,2,3 \, .\label{eq:Z_i}
\end{equation}
The overall renormalisation constant $Z_+$ is then  determined by the condition
(\ref{eq:Z_+}),  using
\begin{equation}
\Gamma^s_+(pa)= \Tr \Pr_0 \Lambda^s_+
            =\left(D_{00}+\sum_{i=1}^3 Z_i D_{i0}\right).
\end{equation}

In eq.\ (\ref{eq:Z_+}),
the renormalisation constant $Z_{\psi}$ is defined by the relation
\begin{equation}
Z_{\psi}(\mu a, g_0^2(a))=
\left.\frac{1}{48}\Tr \left( \Lambda_{V^L_{\mu}}\gamma_{\mu}\right)
              \right|_{p^2=\mu^2}\times Z_{V^L},             \label{eq:Z_psi}
\end{equation}
where $V^L=\bar\psi \gamma_{\mu}\psi$ is the local vector current, and
$Z_{V^L}$ its renormalisation constant which can be determined
with high accuracy, by using  the vector current Ward identities
\cite{wi,wiukqcd,mm}. $\Lambda_{V^L_{\mu}}$ is defined as
\begin{equation}
\Lambda_{V^L_{\mu}}(p)=S(p)^{-1}G_{V^L_{\mu}}(p)S(p)^{-1}\, ,
\end{equation}
where $G_{V^L_{\mu}}(p)$ is the non-amputated two-point Green function
of the local vector current, $G_{V^L_{\mu}}(p)=\langle
\Gamma^{V^L_{\mu}}(p)\rangle$, cf. eq.\ (\ref{eq:gammaa}). There are
various equivalent ways to define $Z_{\psi}$, but (\ref{eq:Z_psi}) is
the most natural from a non-perturbative point of view.  For a more
thorough discussion on the determination of $Z_{\psi}$, we refer the
reader to sections 2 and 4 of  ref. \cite{NP}.

\section{Lattice perturbation theory}
\label{sec:PT}

We have also calculated the renormalisation constants $Z_+, Z_1, Z_2$
and $Z_3$ in one-loop perturbation theory, in order to be able to
compare the results with those obtained non-perturbatively. Since the
non-perturbative renormalisation condition depends on the gauge and on
the external states, the perturbative calculation must be done in the
Landau gauge and at equal external momenta. This calculation is an
extension of those of refs. \cite{4f,improved}.

Starting from a bare lattice operator $O(a)$, the one-loop vertex
function $\Gamma_O^{\lambda}(pa)$ is  obtained by tracing the amputated
Green function
(but with wave function effects included) with a suitable projector.
The generic expression of
$\Gamma_O^{\lambda}(pa)$, calculated between states of
momentum $p$ and in a fixed gauge $\lambda$, is
\begin{equation}
\Gamma_O^{\lambda}(pa)=
\left[1+\frac{\alpha_s}{4\pi}\left(\gamma_O\log(1/p^2 a^2)+
r_O^{\latt}(\lambda,p)
\right)\right] \, .
\end{equation}
$\gamma_O$ is the anomalous dimension, which at one-loop order is
independent of the gauge, the external states and the
regularization\footnote{ $\lambda$ denotes a generic covariant gauge:
$\lambda=0$ corresponds to the Landau gauge,
$\lambda=1$ correspond to the Feynman  gauge.}.
The finite coefficient $r_O^{\latt}(\lambda,p)$, on the other hand, does
depend on  the gauge, the regularization and the external states.
The momentum label $p$ appearing as argument of
$r_O^{\latt}(\lambda,p)$ indicates that the result is a dimensionless
function of  the external states.

In the continuum, in any renormalisation scheme based on dimensional
regularization (DR=NDR, HV or DRED),
the vertex function between states of momentum $p$
and in a generic gauge, is
\begin{equation}
\Gamma_O^{\lambda}\left(p/\mu\right)=
\left[1+\frac{\alpha_s}{4\pi}\left(\gamma_O\log(\mu^2/p^2)+r_O^{\DR}(\lambda,p)
\right)\right]\, ,
\end{equation}
where $\mu$ is the DR renormalisation scale. Thus, the one-loop
relation between the operators in the continuum and on the lattice is
\begin{equation}
 O(\mu)=\left[1+\frac{\alpha_s}{4\pi}
\left(\gamma_O\log(\mu^2a^2)+\Delta^{\DR-\latt}\right)\right] O(a),
\end{equation}
where
\begin{equation}
\Delta_O^{\DR-\latt}=r_O^{\DR}(\lambda,p)-r_O^{\latt}(\lambda,p)
                                      \label{eq:Delta_DR_Latt}
\end{equation}
is independent of both $\lambda$ and $p$.  From
$\Delta_O^{\DR-\latt}$, we can calculate the lattice constant
$r_O^{\latt}(\lambda,p)$, in any gauge and at any external momenta,
from the corresponding constant in the continuum,
$r_O^{\DR}(\lambda,p)$. In order to compare the perturbative result
with the non-perturbative  determination, we need
$r_O^{\latt}(\lambda=0,p)$  in the Landau gauge and with
non-zero but equal external momenta. From (\ref{eq:Delta_DR_Latt}), we
immediately obtain
\begin{equation}
r_O^{\latt}(\lambda=0,p)=r_O^{\DR}(\lambda=0,p)
                              -\Delta_O^{\DR-\latt}.
\end{equation}
Since $r_O^{\latt}$ must be independent of the continuum regularization used
in the intermediate steps,
a check of the correctness of the calculation is given by
\begin{equation}
r_O^{\DRED}(\lambda=0,p)-\Delta_O^{\DRED-\latt}
=r_O^{\NDR}(\lambda=0,p)-\Delta_O^{\NDR-\latt},
\end{equation}
which is equivalent to
\begin{equation}
r_O^{\DRED}(\lambda=0,p)-r_O^{\NDR}(\lambda=0,p) =
r_O^{\DRED}(\lambda=1,p^\prime )-r_O^{\NDR}(\lambda=1,p^\prime ).
\end{equation}

The one-loop contribution to the renormalisation constant $Z_+$  and to
the mixing coefficients $Z_i$'s have been calculated in
\cite{4f,improved}, by comparing the lattice and the  DRED scheme.   In
the notation of these authors
\begin{eqnarray}
Z_+&=&1 + \frac{\alpha_s}{4 \pi} F_+,\qquad F_+=\Delta_{O_+}^{\DRED-\latt}
=-10.9\, ,\nonumber \\
Z_1&=&Z_2=Z_3=\frac{\alpha_s}{4 \pi} F^*,  \qquad F^{\ast}=19.4\, .
\label{contlatt2}
\end{eqnarray}
In the DRED scheme, for a generic gauge $\lambda$ and by  taking the
momenta of the external legs to be equal, we find
\begin{equation}
r_{O_+}^{\DRED}(\lambda,p)=\lambda \Bigl(  - 7/3 + 8/3\log(2) \Bigr)
                          - 5/3 + 8\log(2) \, ,        \label{eq:r_+_DRED}
\end{equation}
whilst the mixing with the ``effervescent" operators is
cancelled by the minimal subtraction procedure.
Thus, the perturbative
expressions of the lattice renormalisation constants
in the RI scheme are
given by\footnote{ The scale $\mu$ in this formula denotes the
renormalization scale  at which $\Gamma_O^{\lambda}(pa)$ is renormalized,
i.e. $Z_+ \Gamma_O^{\lambda}(pa)\vert_{p^2=\mu^2}=1$.}
\begin{eqnarray}
Z_+^{{\rm RI}}&=&1-\frac{\alpha_s}{4\pi}\left(-\gamma_{O_+}\log(\mu^2a^2)
                    +r_{O_+}^{\DRED}(\lambda=0,p)-F_+\right),
\nonumber \\ Z_i^{{\rm RI}}&=&\frac{\alpha_s}{4\pi}F^{\ast},
\end{eqnarray}
with  $\gamma_{O_+}= - 2$  and  $r_{O_+}^{\DRED}(\lambda,p)$ given in
eq.\ (\ref{eq:r_+_DRED}).

We have not evaluated the renormalisation constants using Discrete
Perturbation Theory (DPT), i.e. by summing only over the discrete
values of momenta allowed on our finite lattice (this was done for the
two-quark operators in ref. \cite{NP}). We have only evaluated the
constants using standard lattice perturbation theory, in which finite
lattice size effects are neglected.

In order to estimate the values of the renormalisation constants we
have used the following "boosted" coupling constant $\alpha_S^V$
\cite{Lepage}
\begin{equation}
\alpha_s^V=\frac{1}{\<\frac{1}{3}\Tr U_P\>}\alpha_s^{\latt}
          \simeq 1.68\  \alpha_s^{\latt}
\ \ (\mbox{at} \,\, \beta=6.0)\, .
\end{equation}
as our expansion parameter and refer to the result as corresponding
to Boosted Perturbation Theory (BPT). We also present the values obtained
using the bare coupling $\alpha_s^{\latt}$, and refer to these
results as coming from standard perturbation theory (SPT).

\section{Numerical results}
\label{sec:numerical}
\begin{figure}
\vspace{9pt}
\begin{center}\setlength{\unitlength}{1mm}
\begin{picture}(160,100)
\put(10,-35)
{\includegraphics{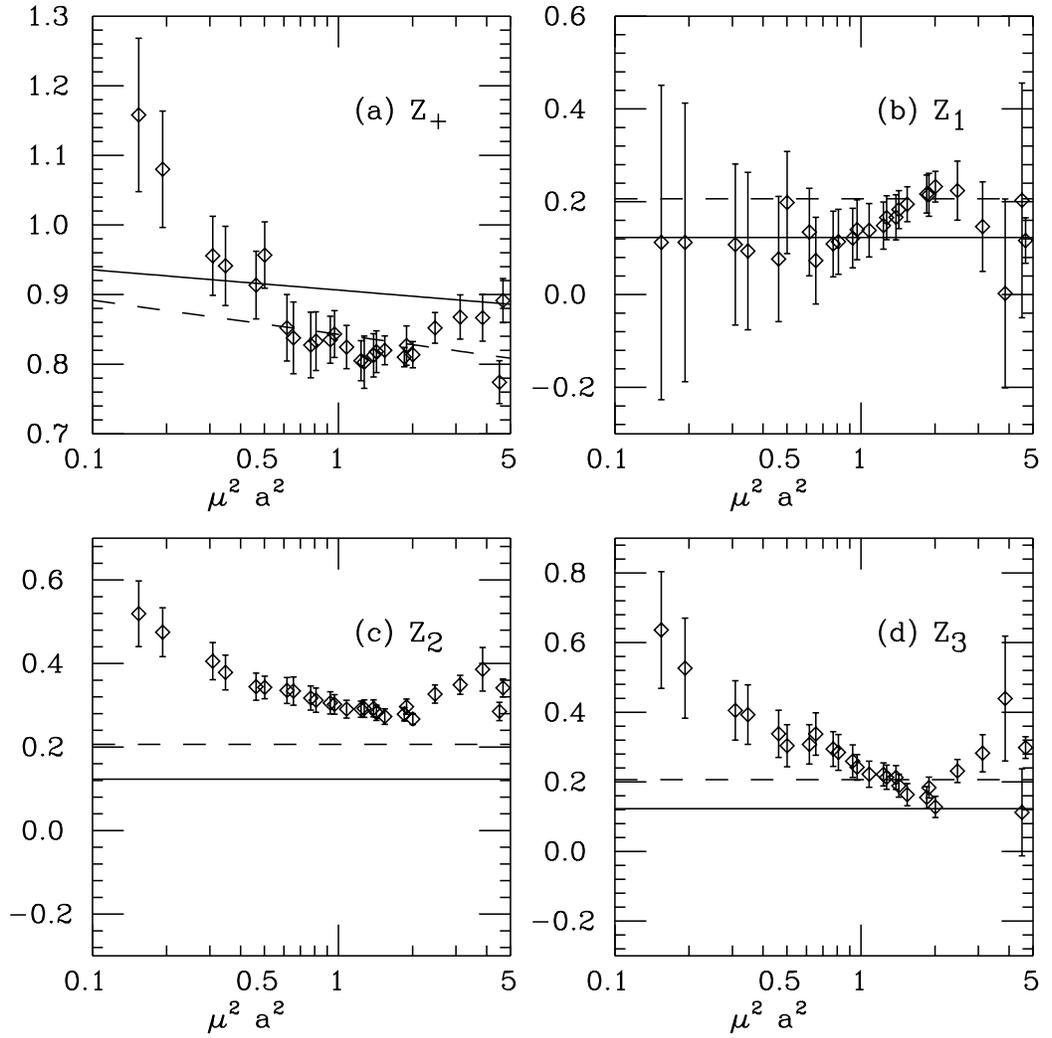}}
\end{picture}
\end{center}

\caption{\it Non-perturbative renormalisation constants of the operator
$O_+$ as a function of $\mu^2 a^2$:  (a)
the overall renormalisation constant $Z_+$; (b)--(d)
the mixing coefficients $Z_i,\ i=1,2,3$.  We also report the
perturbative evaluation:  the dashed curve is from BPT, while the solid
curve is from SPT.}
\label{fig:Z's}
\end{figure}

In this section, we give  the numerical results of our calculation. The
simulation has been performed by generating 36 independent gluon-field
configurations, on a $16^3\times 32$ lattice, at $\beta=6.0$.   The
errors have been obtained with the jacknife method, by decimating three
configurations at a time. The SW-Clover quark propagators have been
computed at a single value of the quark mass  $(m_qa\simeq 0.07)$,
corresponding to the  hopping parameter $\kappa=0.1425$. The quark
Green functions have been computed in the lattice Landau gauge,
defined  by minimizing the functional
\begin{equation}
\Tr \left[ \sum_{\mu=1}^4(U_{\mu}(x)+U_{\mu}^{\dag}(x))\right].
\end{equation}
Possible effects from Gribov copies have not been studied. For more
details, see ref. \cite{paciello}.

In fig. \ref{fig:Z's}, the renormalisation constants, obtained by
using the prescription described in sections \ref{sec:strategy} and
\ref{sec:mixing}, are given as a function of the renormalisation scale
$\mu^2 a^2$.  We hope to find an interval of values of $\mu^2 a^2$,
large enough to avoid significant non-perturbative effects and small
enough to avoid large discretization errors. In ref. \cite{NP}, the
existence of such  a ``window" in $\mu^2 a^2$ was investigated by
comparing the renormalisation constants  of two-quark operators
computed in perturbation theory  with the corresponding
non-perturbative determinations on quark states, and with the results
obtained  by using the Ward identity method \cite{wi,wiukqcd,mm}. In
most of the cases a range  of acceptable values was found in  the
interval  $0.8$--$0.9 \le \mu^2 a^2 \le 1.5$--$2.0$. At smaller values
of $\mu^2 a^2$, in particular in the case of the axial current and of
the pseudoscalar density (probably because of the presence of a
pseudo-Goldstone boson contribution),  the non-perturbative corrections
were found to be large.  For this reason, it is difficult  to determine
the renormalisation constant of the axial current in this way. Only at
values of $\mu^2 a^2$ larger than $1.5$--$2.0$,  a surprisingly large
value in our opinion, did discretization errors become clearly visible.
They were signalled by the fact that the  renormalisation constants
computed at the same values of $\mu^2 a^2$, but with inequivalent
components of the momentum $p$ (e.g. $p\equiv 2 \pi/16 a (4,4,0,2)$ and
$p\equiv 2 \pi/16 a (0,0,0,6)$) were found to be different \cite{NP}.
This was interpreted as a signal of the breaking of the Lorentz
symmetry due to lattice artefacts, see also ref. \cite{nico}.
\begin{table}
\centering
\begin{tabular}{|c|c|c|c|c|}
\hline
$\mu^2 a^2$ &$Z_+$ & $Z_1$ & $Z_2$& $Z_3$ \\
\hline \hline
$0.46$ & $0.91 \pm 0.05$ & $0.08 \pm 0.14$ & $0.34 \pm 0.03$ & $0.34 \pm 0.07$
\\
$0.66$ & $0.84 \pm 0.05$ & $0.07 \pm 0.09$ & $0.33 \pm 0.03$ & $0.34 \pm 0.06$
\\
$0.81$ & $0.83 \pm 0.04$ & $0.11 \pm 0.07$ & $0.31 \pm 0.03$ & $0.28 \pm 0.05$
\\
$0.96$ & $0.84 \pm 0.03$ & $0.14 \pm 0.07$ & $0.30 \pm 0.02$ & $0.24 \pm 0.04$
\\
$1.27$ & $0.80 \pm 0.04$ & $0.17 \pm 0.05$ & $0.29 \pm 0.02$ & $0.21 \pm 0.03$
\\
$1.54$ & $0.82 \pm 0.02$ & $0.19 \pm 0.04$ & $0.27 \pm 0.02$ & $0.16 \pm 0.03$
\\
$1.89$ & $0.83 \pm 0.03$ & $0.22 \pm 0.05$ & $0.30 \pm 0.02$ & $0.18 \pm 0.03$
\\
$2.47$ & $0.85 \pm 0.02$ & $0.22 \pm 0.06$ & $0.33 \pm 0.02$ & $0.23 \pm 0.03$
\\ \hline
SPT    & $0.91$ & $0.12$ & $0.12$ & $0.12$
\\
BPT    & $0.84$ & $0.21$ & $0.21$ & $0.21$
\\ \hline
\end{tabular}
\caption{\it{Values of $Z_+$ and $Z_i$ ($i=1,2,3$) for several
  renormalisation scales $\mu^2 a^2$. We also give the results
obtained at $\mu^2 a^2=1$, by using ``standard"  perturbation theory
(SPT) and ``boosted" perturbation theory (BPT), using an effective
coupling $\alpha_s^V=1.68 \,  \alpha_s^{{\rm latt}}$.}}
\label{tab:examples}
\end{table}

For the four-fermion  operators considered in this paper, we do not
have the possibility of checking the results for the renormalisation
constants by the use of Ward identities, but it appears that a similar
situation may also occur in this case.  In the region of momenta $\mu^2
a^2 \ge  0.96$,  the renormalisation constants are determined  with a
relatively small error (the worse case being the error of $Z_1$ which
is about $50 \%$ at $\mu^2 a^2 = 0.96$) and  the dependence on the
scale is relatively  weak, as  can be seen from  table
\ref{tab:examples} and  fig. \ref{fig:Z's}. We notice that the values
of $Z_+$ and $Z_2$  have small statistical errors even at scales
smaller than $0.96$ ($\sim 10 \%$ in the worst case),
that $Z_3$ has relative errors in the range of about 15-20\%,
and that $Z_1$ has the largest relative error at all the scales
considered in table \ref{tab:examples}. As for the scale dependence,
$Z_2$ is quite stable as a function of $\mu^2 a^2$, while both $Z_1$,
which suffers from the largest statistical uncertainty, and $Z_3$ do
not exhibit a very clear plateau, as can be seen in fig. \ref{fig:Z's}.
In particular the value of $Z_1$ seems to increase
with the scale.  With the present statistics, we cannot determine
whether the variation  of $Z_1$ and $Z_3$ with $\mu^2 a^2$ is real or
due to statistical fluctuations. Fortunately, as we will see below, the
largest correction to the chiral behaviour comes from the operator
$O_2$, corresponding to $Z_2$, which is very well determined.
Hence the chiral behaviour of the operator $O_+$ is
stable with respect to the uncertainties above.

In order to investigate the effects of the non-perturbative
corrections, we have combined our results with the computation of the
lattice matrix elements of the four-fermion operators $O_0$-$O_3$
(\ref{eq:O_+})--(\ref{eq:O_+^SPT}) performed in ref. \cite{Crisafulli},
where a more detailed discussion of the numerical aspects can be found.
Here we limit ourselves to a qualitative discussion of the results.

\begin{figure}
\vspace{9pt}
\begin{center}\setlength{\unitlength}{1mm}
\begin{picture}(160,100)
\put(-10,-60){\includegraphics{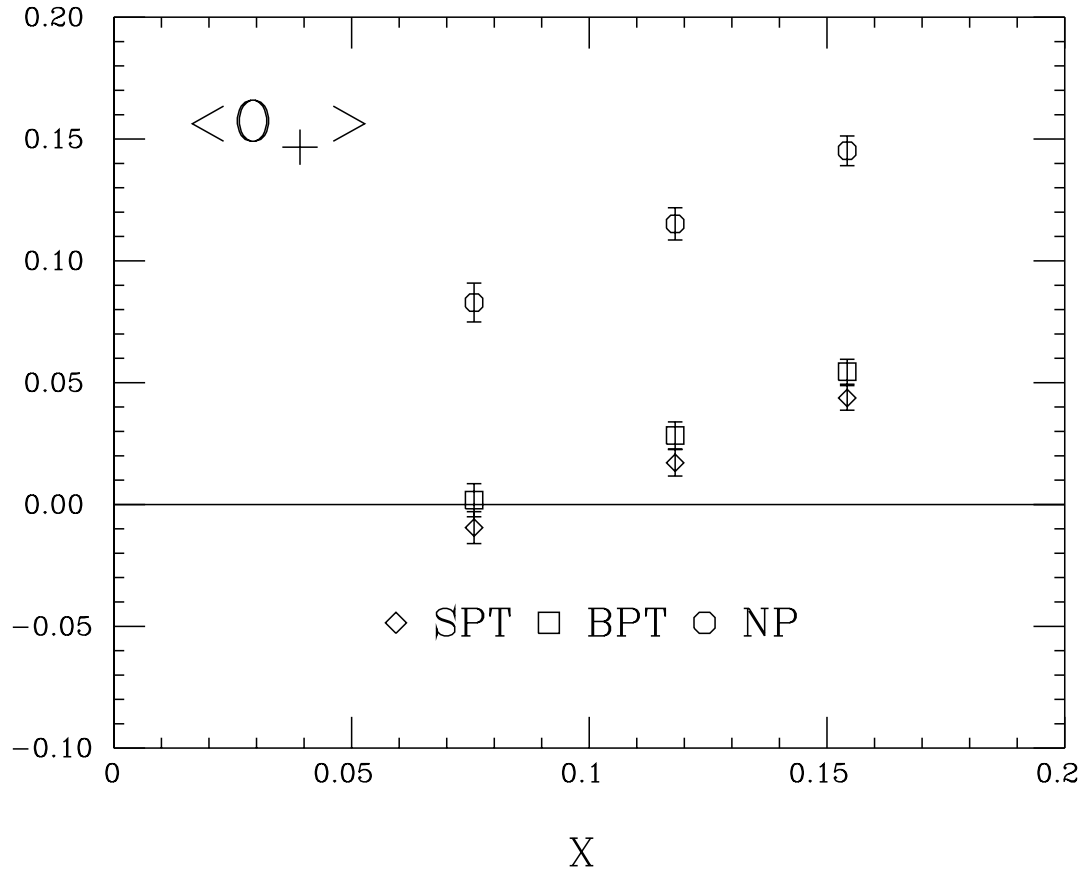}}
\end{picture}
\end{center}

\caption{\it
 Chiral behaviour of  $\<O_+\>$
as a function of $X$ (see text).  We give the
matrix elements of the operator renormalised  in standard
perturbation theory ($\Diamond$), boosted perturbation theory  ($\Box$)
and non-perturbatively (\protect\cerchio).} \label{fig:chiral}
\end{figure}

 In fig. \ref{fig:chiral}, we show the chiral behaviour of
 $\< O_+ \>=\<\bar K^0|
O^{\Delta S=2} | K^0\>_{\latt}/\langle P_5 \rangle^2$
with the meson at rest
  as a function of $X=8/3 f_K^2 M_K^2 /\langle P_5 \rangle^2$.
$\langle P_5 \rangle^2=\vert\langle 0 \vert  \bar s \gamma_5
d \vert K^0 \rangle \vert^2$ is the squared
matrix element of the pseudoscalar density between the meson and the
vacuum. The variables $\< O_+ \>$ and $X$
are  particularly
convenient since they  can be obtained from suitable
two- and three-point  correlation
functions without any fitting procedure \cite{Gavela}.

An analysis of the different
contributions to the matrix element of the renormalised  operator shows
that the largest correction  comes from the  operator $O_2$, eq.\
(\ref{eq:O_+^VA}), whose constant $Z_2$ is well determined. This
contribution is much larger than that coming from $O_1$ (and
larger than that from $O_3$) \footnote{Notice that $O_1$ has
the smallest colour factor.}.
\begin{table}
\centering
\begin{tabular}{|c|c|c|c|c|}
\hline
$\mu^2 a^2$ & $\alpha$ & $\beta$& $\gamma$ \\ \hline \hline
$0.46$ & $ 0.022(16) $ & $0.23(19) $ & $ 0.78(13) $\\
$0.66$ & $ 0.013(15) $ & $0.20(17) $ & $ 0.70(12) $\\
$0.81$ & $ 0.012(13) $ & $0.21(17) $ & $ 0.69(12) $\\
$0.96$ & $ 0.017(13) $ & $0.21(17) $ & $ 0.70(12) $\\
$1.27$ & $ 0.015(13) $ & $0.21(16) $ & $ 0.66(11) $\\
$1.54$ & $ 0.018(13) $ & $0.22(16) $ & $ 0.67(12) $\\
$1.89$ & $ 0.023(13) $ & $0.22(16) $ & $ 0.69(12) $\\
$2.47$ & $ 0.022(14) $ & $0.23(17) $ & $ 0.72(12) $\\
\hline
SPT & $ -0.067(12)$ & $0.17(15)$ & $0.62(11)$ \\
BPT & $ -0.054(12)$ & $0.17(15)$ & $0.62(11)$ \\
\hline
\end{tabular}
\caption{\it{ Values of the coefficients
$\alpha$, $\beta$ and $\gamma$  obtained from a linear fit of
$\< O_+ \>=\<\bar K^0| O^{\Delta S=2} | K^0\>_{\latt}/\< P_5\>^2$.
 The results refer to operators
renormalised at different scales $\mu^2 a^2$. Values of the same parameters
 in standard perturbation theory
and in boosted perturbation theory are also reported.}}
\label{abg}
\end{table}
Thus, the uncertainty in $Z_1$ (and partly $Z_3$)  has no significant
consequences for the value of the kaon matrix element (\ref{eq:B_K}).
Indeed, as can be seen from table \ref{abg}, in passing from
$\mu^2 a^2=0.66$ to $\mu^2a^2=2.47$, the central value of $Z_1$
increases by a factor of 3, but there are no large  variations in the
values of $\alpha, \beta$ and $\gamma$, eq.\ (\ref{eq:B_K_latt}), i.e.
the ``physical" results depend rather weakly on the scale.
The use of the non-perturbative renormalisation constants leaves the
values of $\beta$ and $\gamma$ almost unchanged compared to those
obtained by using the constants computed in one-loop perturbation
theory: $\beta \sim 0.2$, and within the errors is compatible with zero
in both cases, and $\gamma$ is about $15 \%$ larger in the
non-perturbative case. In contrast, $\alpha$ changes sign and its
absolute value is reduced by about a factor of three in the
non-perturbative case, and becomes compatible with zero. This happens
for any choice of $\mu^2 a^2$ between $0.46$ and $2.47$, cf. tab.\
\ref{tab:examples}.
The variation of the $Z_i$'s in the  interval of $\mu^2 a^2$ considered
in tables \ref{tab:examples} and \ref{abg} is representative of the
variation allowed by the statistical errors. Since the $Z_i$'s and the
matrix elements of the four-fermion operators have been computed on
different sets of configurations, this is the  best test of the
stability of the results which can be done at present. A more definite
conclusion will be reached by computing the renormalisation constants
and the matrix elements on the same set of configurations.

Since $\alpha$ should vanish in the continuum limit,  we conclude that
the use of the non-perturbative renormalisation constants improves the
chiral behaviour for a large range of values of the renormalisation
scale.
 This is also illustrated in fig. \ref{fig:chiral}
\cite{Crisafulli}, obtained for $\mu^2 a^2=0.96$.
 More details will be given in ref.\ \cite{Crisafulli}.

\section{Conclusions}
\label{sec:conclusion}
We have applied the non-perturbative renormalisation method proposed in
ref. \cite{NP} to the $\Delta S=2$ operator given in eq.\
(\ref{eq:O_DS=2}). Since, on the lattice, this operator  mixes  with
other dimension-six operators of different chirality, we have
illustrated a projection method for the determination of the mixing
coefficients. The overall renormalisation constant of the subtracted
operator has then been obtained as in the case of any other
multiplicatively renormalisable operator.

In this exploratory study we have computed the subtraction constants
with limited statistical precision (36 configurations on a $16^3\times
32$ lattice at $\beta=6.0$ using the improved SW-Clover action). The
results in fig.\ref{fig:Z's} and table \ref{tab:examples} are very
encouraging, and motivate us to repeat the calculation with larger
statistics and at different values of $\beta$ ($\beta=6.2$ and $6.4$),
and to extend it to the operators relevant for $\Delta I=1/2$
transitions and to the penguin operators which control CP-violation in
kaon systems. Even with our limited statistical precision, our results
indicate that the chiral behaviour of the $K^0$--$\bar K^0$ matrix
element of $O^{\Delta S=2}(\mu)$ is improved significantly by the  use
of the subtraction constants which were determined non-perturbatively.
This supports our view that, by combining the improvement of the action
{\em \`a la} Symanzik, which reduces $O(a)$ effects, with the
non-perturbative method of ref. \cite{NP},  which reduces higher-order
effects in the mixing coefficients, it is possible to achieve an
accurate determination of the physical weak amplitudes using
Wilson-like fermions.

\section*{Acknowledgements}
We warmly thank the members of the APE collaboration for the use of
their results before publication. M.T. thanks the Physics Department of
Southampton University for their kind hospitality during the
completion of this work. We acknowledge the partial support by the EC
contract CHRX-CT92-0051 and by M.U.R.S.T., Italy.  C.T.S. acknowledges
the Particle Physics and Astronomy Research Council for its support
through the award of a Senior Fellowship. We also acknowledge the
Computer Centre of CINECA (Bologna, Italy), where these calculations
were performed, and thank their staff for their precious help.


\begin{thebibliography}{99}
\bibitem{NP}
G. Martinelli, C. Pittori, C.T. Sachrajda, M. Testa and A. Vladikas,
Nucl. Phys. B445 (1995) 81.
\bibitem{Cabibbo}
N. Cabibbo, Phys. Rev. Lett. 12 (1964) 62.
\bibitem{Gellmann}
M. Gell-Mann, Phys. Rev. Lett. 12 (1964) 155.
\bibitem{sw}
B. Sheikholeslami and R. Wohlert, Nucl. Phys. B259 (1985) 572.
\bibitem{marti84}
G. Martinelli, Phys. Lett. B141 (1984) 395.
\bibitem{Bochicchio}
M. Bochicchio, L. Maiani, G. Martinelli, G.C. Rossi and M. Testa,
Nucl. Phys. B262 (1985) 331.
\bibitem{Maiani}
L. Maiani, G. Martinelli, G.C. Rossi and M. Testa,
Phys. Lett. B176 (1986) 445; Nucl. Phys. B289 (1987) 505.
\bibitem{berw}
C. Bernard, T. Draper and A. Soni, Phys. Rev. D36 (1987) 3224.
\bibitem{4f}
R. Frezzotti, E. Gabrielli, C. Pittori and G.C. Rossi,
Nucl. Phys. B373 (1991) 781.
\bibitem{improved}
A. Borrelli, R. Frezzotti, E. Gabrielli and C. Pittori,
Nucl. Phys. B409 (1993) 382.
\bibitem{bern}
C. Bernard, T. Draper, G. Hockney, A.M. Rushton and A. Soni,
Phys. Rev. Lett. 55 (1985) 2770.
\bibitem{capri}
APE collaboration, presented by G. Martinelli,
Nucl. Phys. B (Proc. Suppl.) 17 (1990) 523.
\bibitem{Gavela}
M.B. Gavela {\em et al.}, Nucl. Phys. B306 (1988) 677.
\bibitem{Bernard}
C. Bernard and A. Soni,
Nucl. Phys. B (Proc. Suppl.) 17 (1990) 495;
Nucl. Phys. B (Proc. Suppl.) 20 (1991) 410.
\bibitem{gupta}
R. Gupta, D. Daniel, G.W. Kilcup, A. Patel and S. Sharpe,
Phys. Rev. D46 (1993) 5113.
\bibitem{Donini94}
APE Collaboration, M. Crisafulli {\em et al.}, presented by A. Donini,
Nucl. Phys. { B} (Proc. Suppl.) 42 (1995) 397. \\
T. Bhattacharya and R. Gupta,
Nucl. Phys. { B} (Proc. Suppl.) 42 (1995) 935.
\bibitem{Crisafulli} APE Collaboration,
M. Crisafulli {\em et al.},   ROME Prep. 1105/95, in preparation.
\bibitem{clover}
G. Heatlie, G. Martinelli, C. Pittori, G.C. Rossi and C.T. Sachrajda,
Nucl. Phys. B352 (1991) 266.
\bibitem{msv}
G. Martinelli, C.T. Sachrajda and A. Vladikas, Nucl. Phys. B358 (1991) 212.
\bibitem{tass2}
G. Martinelli, C.T. Sachrajda, G. Salina and A. Vladikas,
Nucl. Phys. B378 (1992) 591.
\bibitem{wi}
G. Martinelli, S. Petrarca, C.T. Sachrajda and A. Vladikas,
Phys. Lett. B311 (1993) 241.
\bibitem{wiukqcd}
UKQCD Collaboration,
D.S. Henty, R.D. Kenway, B.J. Pendleton and J.I. Skullerud,
Edinburgh prep. 94/545, hep-lat 9412088.
\bibitem{Lepage}
G.P. Lepage and P.B. Mackenzie,
Nucl. Phys. B (Proc. Suppl.) 20 (1991) 173; Phys. Rev. D48 (1993) 2250.
\bibitem{Ciuchini2}
M. Ciuchini, E. Franco, G. Martinelli and L. Reina,
Rome prep. 94/1024, CERN-TH 7514/94, hep-ph 9501265, to appear
in Z. Phys. C.
\bibitem{Altarelli}
G. Altarelli, G. Curci, G. Martinelli and S. Petrarca,
Nucl. Phys. B187 (1981) 461.
\bibitem{Buras}
A.J. Buras, M. Jamin, M.E. Lautenbacher and P.H. Weisz,
Nucl. Phys.  B370 (1992) 69; Addendum Nucl Phys. B375 (1992) 501.
\bibitem{Ciuchini}
M. Ciuchini, E. Franco, G. Martinelli and L. Reina,
Nucl. Phys.  B415 (1994) 403.
\bibitem{BERNARD2}
C.~Bernard, T. Draper, G. Hockney and A.~Soni,
Nucl. Phys. B (Proc. Suppl.) 4 (1988) 483.
\bibitem{mm}
L. Maiani and G. Martinelli, Phys. Lett. B178 (1986) 265.
\bibitem{paciello} M.L. Paciello, S. Petrarca, B. Taglienti and
A. Vladikas, Phys. Lett. B341 (1994) 187.
\bibitem{nico} P. Marenzoni, G. Martinelli and N. Stella,
Rome prep. 94/1042- SHEP prep. 93/94-31,
hep-lat 9410011,  to appear in  Nucl. Phys. B.
\end{thebibliography}
\end{document}